\documentclass{PoS}
\usepackage{epsfig}
\usepackage{latexsym}
\usepackage{amsmath}
\usepackage{amsfonts}
\newcommand{\Tr}{\mathop{\textrm{Tr}}}
\renewcommand{\Re}{\mathop{\textrm{Re}}}

\newcommand{\Nc}{N_{\rm c}}
\newcommand{\Tc}{T_{\rm c}}
\newcommand{\msbar}{{\overline{\mbox{\rm MS}}}}
\newcommand{\tinymsbar}{{\overline{\mbox{\tiny\rm{MS}}}}}
\newcommand{\tbG}{\tilde B_\tinymsbar}
\newcommand{\rmi}[1]{{\mbox{\scriptsize #1}}}

\newcommand{\fr}[2]{{\frac{#1}{#2}\,}}

\title{Renormalization of infrared contributions to the QCD pressure}

\ShortTitle{Renormalization of infrared contributions \\ to the QCD pressure}

\author{\speaker{C. Torrero}, M. Laine, Y. Schr\"oder \\

        Faculty of Physics, University of Bielefeld, 
        33501 Bielefeld, Germany\\

        E-mail: 
        \email{torrero@physik.uni-bielefeld.de},

        \email{laine@physik.uni-bielefeld.de},

        \email{yorks@physik.uni-bielefeld.de}}

\author{F. Di Renzo\\

        Università di Parma \& INFN, Parco Area delle Scienze 7A, 
        43100 Parma, Italy\\

        E-mail: \email{direnzo@fis.unipr.it}}

\author{V. Miccio\\

        Università di Milano Bicocca \& INFN,
        Piazza dell'Ateneo Nuovo 1, 20126 Milano, Italy\\

        E-mail: \email{vincenzo.miccio@mib.infn.it}}

\abstract{Thanks to dimensional reduction, the infrared 
contributions to the QCD pressure can be obtained from two different
three-dimensional effective field theories, called the Electrostatic
QCD (Yang-Mills plus adjoint Higgs) and the Magnetostatic QCD (pure
Yang-Mills theory). Lattice measurements have been carried out within
these theories, but a proper interpretation of the results requires
renormalization, and in some cases also improvement, i.e. the removal
of terms of $O(a)$ or $O(a^2)$. We discuss how these computations can
be implemented and carried out up to 4-loop level with the help of
Numerical Stochastic Perturbation Theory.}

\FullConference{XXIVth International Symposium on Lattice Field Theory\\

                July 23-28, 2006\\

                Tucson, Arizona, USA}

\begin{document}

%%%%%%%%%%%%%%%%%%%%%%%%%%%%%%%%%%%%%%%%%%%%%%%%%%%%%%%%%%%%%%%%%%%%%%%%%%%%%
%

\section{Introduction}

As is well-known, the QCD pressure is an observable that plays 
a role in many contexts: besides being important for theoretical
studies of the QCD phase transition and thermodynamics, it also has potential
phenomenological relevance for cosmology and heavy ion collision experiments.

At high temperatures, a useful approach for the computation of this 
observable is \emph{Dimensional Reduction}~\cite{GAP,KLRS,BN}. 
It consists of replacing the full 4d theory with an effective 3d 
one including an adjoint Higgs field
(Electrostatic QCD, ``EQCD''). This theory can, 
in turn, be reduced to a 3d pure Yang-Mills theory
(Magnetostatic QCD, ``MQCD''). This strategy is useful first of all from 
the theoretical point of view, since it allows for a separation of
contributions coming from the various scales that characterize QCD,
namely $T$ (hard modes), $gT$ (soft modes) and $g^2T$ 
(ultrasoft modes). Moreover, it permits a study of the 
whole $T$-range of interest: 
the high-temperature region is usually investigated by means of
perturbation theory while the low-temperature regime is explored via
lattice simulations. There might, however, be a gap between the two regimes:
on the perturbative side it is not possible to lower $T$ too much because 
of the poor convergence~\cite{az}, while on the lattice side
numerical limitations forbid simulations at temperatures higher
than about $5\Tc$~\cite{boyd}. Dimensional Reduction can overlap with both
of these regimes and thus fill the possible gap.
   
Within this framework, our first aim is to complete the determination of 
the order ${O}(g^6)$ weak-coupling expansion of the QCD pressure: due to
the presence of IR divergences~\cite{LINDE}, non-perturbative lattice
measurements are needed at this order~\cite{HKLRS}, 
but their proper interpretation
in the context of the full computation~\cite{gsixg,pheneos} 
requires a conversion of the
regularization scheme from lattice to $\msbar$. Second, the full 
Dimensional Reduction program requires the study of EQCD~\cite{a0cond}, 
but the continuum extrapolations that enter at this stage turn out
to be very delicate, and require the
removal of lattice artifacts at $O(a)$ and $O(a^2)$.

Our aim is to compute these renormalization constants and
improvement coefficients by means of \emph{Numerical Stochastic
Perturbation Theory} (NSPT), a procedure developed by the Parma group.

%%%%%%%%%%%%%%%%%%%%%%%%%%%%%%%%%%%%%%%%%%%%%%%%%%%%%%%%%%%%%%%%%%%%%%%%%%%%
%

\section{NSPT basics}

NSPT has its origins in the concept of \emph{Stochastic
Quantization}~\cite{PW} whose recipe is made up of two ingredients: 
the introduction of an extra coordinate, a stochastic time $t$, 
and an evolution equation of the Langevin type,
\begin{equation}
 \frac{\partial\phi(x,t)}{\partial t}=
  -\frac{\partial S[\phi]}{\partial\phi}+\eta(x,t)~,
\end{equation}       
where $\eta(x,t)$ is a Gaussian noise.
Starting from this, the usual Feynman-Gibbs integration can be
reproduced by averaging over the noise $\eta$, or more 
practically over the stochastic time $t$, that is
\begin{equation}
 Z^{-1}\!\!\int [D\phi]
 O[\phi\small(x\small)]e^{-S[\phi(x)]}=\lim_{t\rightarrow\infty}
 \frac{1}{t} \int_0^t \! {\rm d} t' \,
 \big\langle O[\phi_{\eta}\small(x,t'\small)]\big\rangle_{\eta}\;.
\end{equation}   
When dealing with $SU(3)$ variables, the Langevin equation needs to be
modified into
\begin{equation}
 \partial_{t}U_{\eta} = -i\Bigl( \nabla S[U_{\eta}]+\eta \Bigr)
 U_{\eta} \;,
\end{equation}
in order to assure the correct evolution of the variables 
within the group.

In this framework, perturbation theory can be 
introduced by means of the expansion~\cite{DR1}
\begin{equation}
 U_{\eta}(x,t)\longrightarrow\sum_{k}g_0^kU_{\eta}^{(k)}(x,t) 
 \;,
\end{equation} 
where $g_0$ is the bare gauge coupling. 
This gives a system of coupled differential equations that can be
solved numerically via a discretization of the stochastic time $t=n\tau$, 
where $\tau$ is a time step. In
practice, we let the system evolve according to the Langevin equation
for different values of $\tau$, average over each
thermalized signal (this is the meaning of the above-mentioned limit
$t\rightarrow \infty$), and then extrapolate in order to get the
$ \tau = 0 $ value of the desired observable. This procedure is then
repeated for different values of the various parameters appearing in
the action.

%%%%%%%%%%%%%%%%%%%%%%%%%%%%%%%%%%%%%%%%%%%%%%%%%%%%%%%%%%%%%%%%%%%%%%%%%%%%
%

\section{Renormalization of the Magnetostatic sector: setup}

The MQCD contribution to the pressure can be written as~\cite{gsixg} 
\begin{equation}
 f_{\scriptscriptstyle\overline{MS}}=
 -g_M^6\frac{d_AN_c^3}{(4\pi)^4} 
 \bigg[\bigg(\frac{43}{12}-\frac{157}{768}\pi^2\bigg)
 \ln\frac{\bar{\mu}}{2N_c~\!g_M^2}+B_G+{O}(\epsilon)\bigg]~,
\end{equation} 
with $N_c$ the number of colours, $d_A\equiv N_c^2-1$, 
$\bar\mu$ the $\msbar$ scheme scale parameter, 
$g_M$ the gauge coupling and $B_G$ the constant 
we ultimately want to determine. It can be shown that
for $\Nc = 3$~\cite{OURS}, 
\begin{equation}
 B_G = 10.7 \pm 0.4  - \tilde B_\rmi{L}(1)  + \tbG(1)
 + \biggl( \frac{43}{12} - \frac{157}{768} \pi^2 \biggr)
 \biggl(
  \fr13 + \ln2 + 2 \ln \Nc 
 \biggr)
 \;, \label{master}
\end{equation}
where the numerical value is
non-perturbative and follows from lattice simulations~\cite{HKLRS}
(for the corresponding numbers at $\Nc\neq 3$, see ref.~\cite{kh}), 
while $\tbG(1) = -2.16562591949800919016$ has been determined
as a result of extensive continuum computations~\cite{sun,sv}.
The remaining unknown, $\tilde B_\rmi{L}(1)$, can be expressed 
as~\cite{OURS}
\begin{equation}
 8\frac{d_A\Nc^{~\!\!6}}{(4\pi)^4}
 \widetilde{B}_L(1)=
 \lim_{m\rightarrow0}\beta_0^4
 \bigg\{\left\langle1-\frac{1}{\Nc}\Tr[\widetilde{P}_{12}]
 \right\rangle_\rmi{up~to~4-loop}
 -\bigg[\frac{c_1}{\beta_0}+\frac{c_2}{\beta_0^2}+
 \frac{c_3}{\beta_0^3}+\frac{c_4}{\beta_0^4} 
 \ln\frac{1}{am}\bigg]\bigg\}~,
 \label{master2}
\end{equation}  
where the argument ``1'' corresponds to Feynman gauge, 
$\beta_0\equiv 2\Nc/ag_M^2$,~\!$a$ is the lattice spacing, 
$\Tr[\widetilde{P}_{12}]$ is the trace of the elementary 
plaquette in the 1-2 plane, and $m$ is a gluon mass the has
been introduced as an intermediate IR regulator. 
The coefficients $c_1,...,c_4$
are all known~\cite{hk,PRyork,pt,HKLRS}. 

To evaluate eq.~(\ref{master2}), gauge fixing and mass terms 
need to be introduced: 
\begin{equation}
 Z=
 %\int\! [D\phi] \,
 %{\Det\Bigl(-\sum_{\mu} \hat{\partial}_{\mu}^{L}\hat{D}_{\mu}[\phi]+m^2\Bigr)}
 %\exp\big(-S_{\scriptscriptstyle{W}}-S_{\scriptscriptstyle{GF}}\big)=
 \int\! [D\phi]\,
 \exp\big(-S_{\scriptscriptstyle{W}}-S_{\scriptscriptstyle{GF}}-
  S_{\scriptscriptstyle{FP}}\big)\;,
\end{equation}
where we assume the use of lattice units (i.e. $a=1$), and 
\begin{eqnarray}
 S_{\scriptscriptstyle{W}} 
 & = & \beta_{0}\sum_{P}(1-\Pi_{P})
 +\frac{\beta_{0}m^2}{4\Nc}\sum_{x,\mu,A}\phi_{\mu}^{A}(x)\phi_{\mu}^{A}(x)
 \;, \\
%\end{equation}
%\begin{equation}
 S_{\scriptscriptstyle{GF}} 
 & = & \frac{\beta_{0}}{4\Nc} %% {4\Nc\alpha}
 \sum_{x,A}\Bigl[\sum_{\mu}
 \hat{\partial}_{\mu}^{L}\phi_{\mu}^{A}(x)\Bigr]^2 
 \;, \\
%\end{equation}
%\begin{equation}
 S_{\scriptscriptstyle{FP}} 
 & = & -\Tr\Bigl[\ln\Bigl(-
 \sum_{\mu}\hat{\partial}_{\mu}^{L}\hat{D}_{\mu}[\phi]+m^2\Bigr)\Bigr]
 \;.
\end{eqnarray}   
Here we have followed the conventions of ref.~\cite{ROTHE}, 
writing in particular 
$U_\mu = \exp(i \phi_\mu)$, 
$\phi_\mu = \phi_\mu^A T^A$,
with the normalization $\Tr[T^AT^B] = \delta^{AB}/2$.
Moreover $m$ is the common gluon and ghost mass and 
$\hat D_{\mu}$ is the discrete Faddeev-Popov operator, 
given by~\cite{ROTHE}
\begin{equation}
 \hat{D}_{\mu}[\phi]=
 \biggl[1+\frac{i}{2}\displaystyle \Phi_{\mu}
 -\frac{1}{12} \displaystyle \Phi_{\mu}^2
 -\frac{1}{720} \displaystyle \Phi_{\mu}^4
 -\frac{1}{30240}\displaystyle \Phi_{\mu}^6
 +{O}(\displaystyle \Phi_{\mu}^8)\biggr]
  \hat{\partial}_{\mu}^{R}+i\displaystyle \Phi_{\mu}
 \;,
\end{equation}
with $\displaystyle \Phi_\mu = \phi_\mu^A F^A$, 
where $[F^A]_{BC} \equiv - i f^{ABC}$
are the generators of the adjoint representation. 
Details about the treatment of the Faddeev-Popov 
determinant can be found in ref.~\cite{OURS}.

The procedure then consists of measuring the plaquette for different
lattice sizes at fixed mass, then extrapolating towards infinite
volume, and repeating for different values of the mass. 
Finally, after subtracting the logarithmic divergence, the
zero-mass extrapolation of eq.~(\ref{master2})
will provide the quantity we want to measure. It
is important to perform first the extrapolation in volume and then in
mass, because the opposite order would result in having the finite size
as the IR regulator and not the mass as desired.

%%%%%%%%%%%%%%%%%%%%%%%%%%%%%%%%%%%%%%%%%%%%%%%%%%%%%%%%%%%%%%%%%%%%%%%%%%%%
%

\section{Renormalization of the Magnetostatic sector: results}

As just stated, the first step is the extrapolation in volume. 
While the analytic behavior is known at 1-loop level (and this has 
given us a useful crosscheck), this 
is not true at 4-loop level, so that we have to rely on effective fits. 
Some of them are shown in Fig. 1.

In order to be as conservative as possible, we opted for fitting a
constant to those points that do not seem to show any volume
dependence within errorbars. To check whether this approach is
reliable, we employed it for the first three loops, and then
performed the zero-mass extrapolation to see whether the already known
coefficients $c_1$,~$c_2$ and $c_3$ are recovered.
Table 1 confirms that this indeed is the case. 

%
%%%%%%%%%%%%%%%%%%%%%%%%%%%%%%%%%%%%%%%%%%%%%%%%%%%%%%%%%%%%%%%%%%%%%%%%
%
\begin{table}[h]
\begin{center}
\begin{tabular}{|r|r|r|}
\hline
Coefficient & Our extrapolation & Known result \\
\hline
$c_1$ & 2.672(8) &  2.667 \\
\hline
$c_2$ & 1.955(16) &  1.951 \\
\hline
$c_3$ & 6.83(10)  &  6.86 \\
\hline
\end{tabular}
\caption{Comparison of our extrapolations 
with the known results~\cite{hk,pt,HKLRS} for the first three loops.}
\label{Tab.1}
\end{center}
\end{table}
%%%%%%%%%%%%%%%%%%%%%%%%%%%%%%%%%%%%%%%%%%%%%%%%%%%%%%%%%%%%%%%%%%%%%%%%%%%%%

The same procedure was subsequently applied 
at the 4-loop order where, however, the IR divergence
needs to be subtracted before taking the zero-mass limit 
(see Fig.~2). We then performed polynomial extrapolations 
involving a different number of points and degrees of freedom: 
the final results we get for $\widetilde{B}_L(1)$ is~\cite{OURS}
\begin{equation}
 \widetilde{B}_L(1)=13.8\pm0.4~,
\end{equation} 
which gives, once inserted into eq.~(\ref{master}),
\begin{equation}
 B_G=-0.2\pm0.4^{(MC)}\pm0.4^{(NSPT)}~.
\end{equation} 
Here ``MC'' labels the result of Monte Carlo simulations~\cite{HKLRS}.

%%%%%%%%%%%%%%%%%%%%%%%%%%%%%%%%%%%%%%%%%%%%%%%%%%%%%%%%%%%%%%%%%%%%%%%%%%%%
%  
\begin{figure}[t]
% \hfill
 \begin{minipage}[t]{.45\textwidth}
 %\begin{center} 
   \hspace*{-0.5cm}%
   \epsfig{file=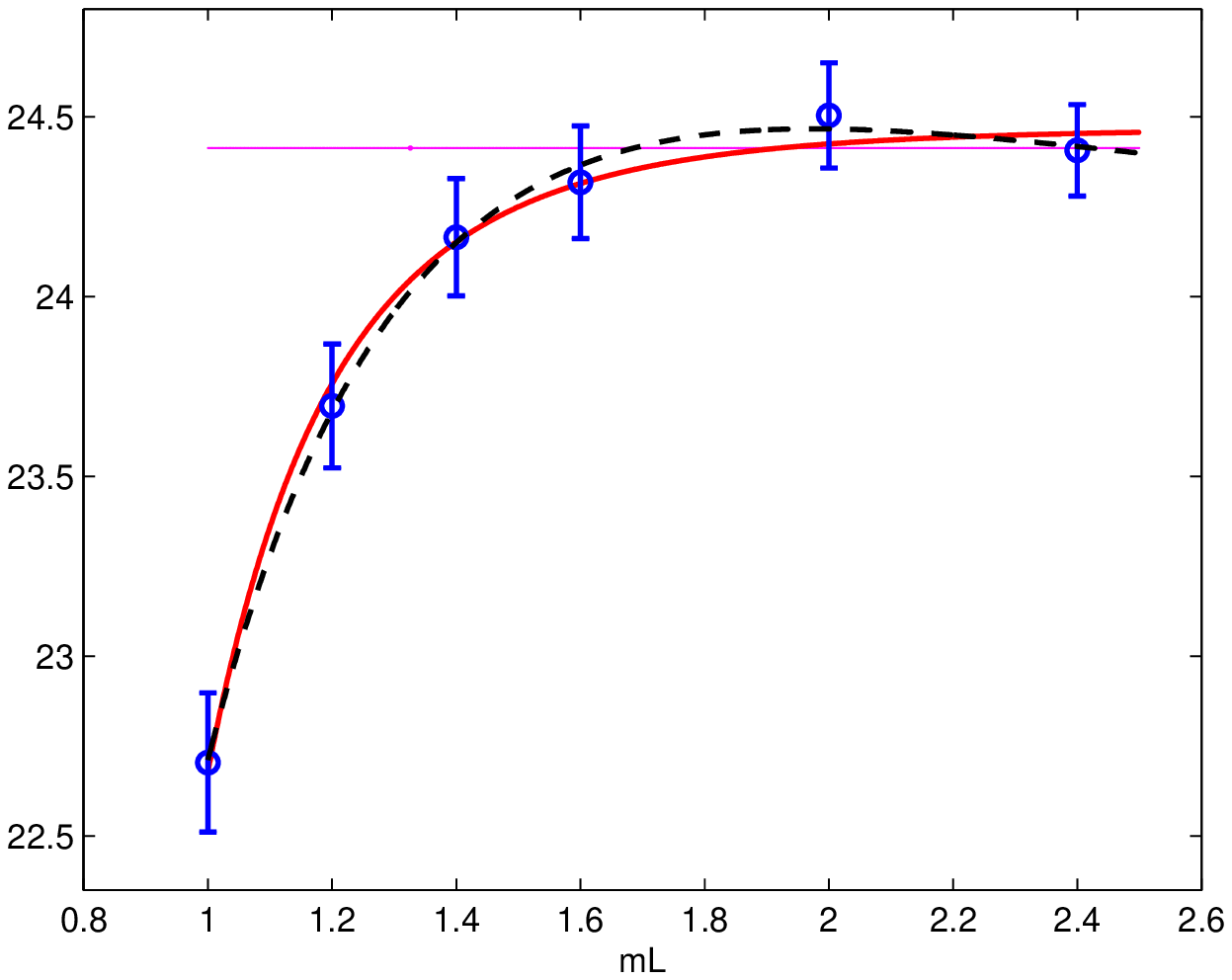, scale=0.55}
   \caption{4-loop plaquette vs $mL$ at fixed mass ($m=0.2$): 
   the solid and the dashed lines correspond to different combinations 
   of a constant plus a sum of negative exponentials and negative powers 
   of $mL$; the horizontal line fits a constant.}
   \label{Fig.1}
  %\end{center}
 \end{minipage}
 \hfill
 \begin{minipage}[t]{.45\textwidth}
  %\begin{center} 
   \epsfig{file=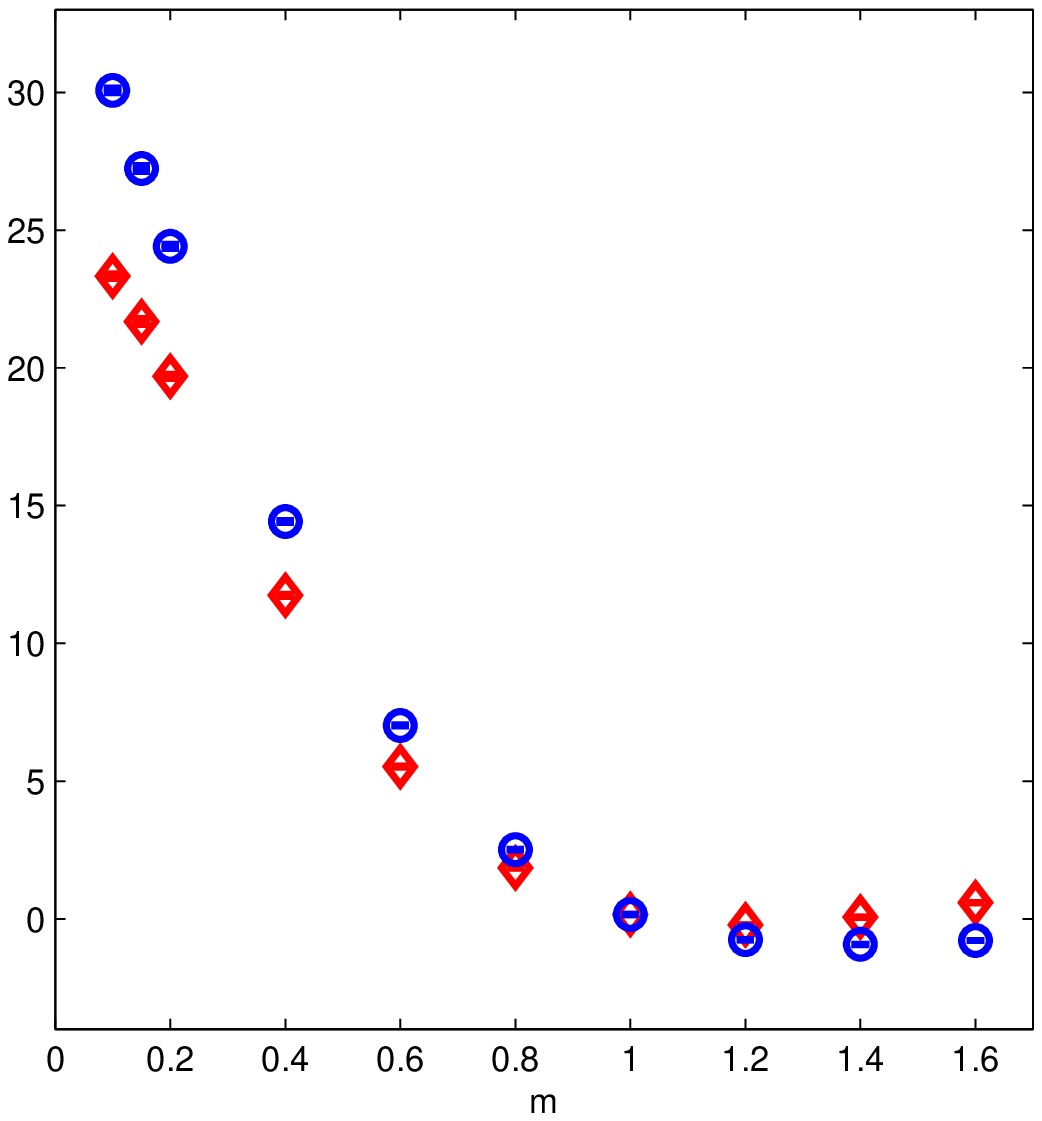, scale=0.50}
   \caption[a]{Infinite-volume 4-loop plaquette vs $m$: 
   circles still contain $\ln(m)$ while diamonds are obtained 
   after its subtraction (cf.~eq.~(\ref{master2})).}
   \label{Fig.2}
  %\end{center}
 \end{minipage}
 \hfill
\end{figure}
%
%%%%%%%%%%%%%%%%%%%%%%%%%%%%%%%%%%%%%%%%%%%%%%%%%%%%%%%%%%%%%%%%%%%%%%%%%%%

%%%%%%%%%%%%%%%%%%%%%%%%%%%%%%%%%%%%%%%%%%%%%%%%%%%%%%%%%%%%%%%%%%%%%%%%%%%%
%

\section{Renormalization and improvement in the Electrostatic sector: setup}

The EQCD action in the continuum is given by
\begin{equation}
 S_E=\int \! {\rm d}^dx \,
 \biggl\{\frac{1}{2}\Tr[F_{ij}^2(x)]+\Tr[D_i,A_0(x)]^2
 +m^2\Tr [A_0^2(x)]+\lambda\big(\Tr[A_0^2(x)]\big)^{\!2}
 \biggr\}~,
\end{equation}
where $F_{ij}(x)$ is the 3d field strength tensor, 
$D_i$ the covariant derivative and $A_0=\sum_{B=1}^8A_0^BT^B$,
with $T^B$  normalised as before.
Once again the strategy is to obtain
the $\msbar$ scheme result starting from lattice
measurements. Apart from the plaquette expectation value, 
this theory has however more  \emph{condensates} that play a role. 
In particular, derivatives with respect to 
the dimensionless variables $y$ and $x$, defined as
% \begin{equation}
% y=\frac{m^2}{g_E^4}~,~~~~~~~~~x=\frac{\lambda}{g_E^2}~,
% \end{equation}
$
 y = m^2/g_E^4
$, 
$ 
 x = \lambda / g_E^2
$
(with $g_E$ the EQCD gauge coupling), 
produce condensates quadratic and quartic in $A_0$~\cite{a0cond}.
Subtracting the proper counterterms~\cite{contlatt}, 
and ${O}(a)$ or ${O}(a^2)$ effects, which 
become important in the range of large $y$ where connection
to the weak-coupling expansion can be made, we can extrapolate to the
continuum, and finally obtain the pressure by integration~\cite{a0cond}.

The EQCD lattice action is given by
\begin{equation}
 \nolinebreak
 \begin{array}{ccl}
 S_\rmi{latt}&=&\beta\sum_{x,~\!i<j}
 \Bigl(1-\frac{1}{3}\Re\Tr\big[P_{ij}(x)\big]\Bigr)
 -2\sum_{x,~\!i}
 \Tr\big[\phi(x)~\!U_i(x)~\!\phi(x+i)~\!U_i^{\dag}(x)\big]+\\
 &+&\sum_x\bigg\{\alpha(\beta,\lambda,y_\rmi{latt}) 
 \Tr\big[\phi^2(x)\big]
 +\lambda\Big(\!\Tr\big[\phi^2(x)\big]\Big)^2 \bigg\}~,
\end{array}
\end{equation}
where now $\beta=2\Nc/ag_E^2$, %%~\! $P_{ij}$ is the lattice plaquette, 
$U_i$ is the link variable, $\phi=A_0\sqrt{6/\beta}$,  and~\cite{contlatt} 
\begin{equation}
\nolinebreak
\begin{array}{ccl}
 \alpha(\beta,\lambda,y_\rmi{latt})&=&
 6\bigg\{1+\frac{1}{6}y_\rmi{latt}-(6+\frac{5}{3}\lambda\beta) 
 \frac{3.175911525625}{4\pi\beta}-\\
 &-&\frac{3}{8\pi^2\beta^2}
 \Big[(10\lambda\beta-\frac{5}{9}\lambda^2\beta^2)(\ln\beta+0.08849)+
 \frac{34.768}{6}\lambda\beta+36.130\Big]\bigg\}~.
\end{array}
\end{equation}

The first condensate we want to measure is the derivative 
with respect of $y_\rmi{latt}$: apart from a rescaling factor, 
it is equal to $\langle \Tr[A_0^2]\rangle$ whose 
lattice counterpart has a perturbative expansion given by
\begin{equation}
\begin{array}{ccl}
 \langle~\! \Tr~\![~\!\phi^2]~\!\rangle&=&
 d_{00}+d_{10}\frac{1}{\beta}+d_{11}\lambda+
 d_{20}\frac{1}{\beta{^2}}+d_{21}\frac{\lambda}{\beta}+d_{22}\lambda^2+\\
 &+&d_{30}\frac{1}{\beta{^3}}+
    d_{31}\frac{\lambda}{\beta{^2}}+ 
    d_{32}\frac{\lambda^2}{\beta}+
    d_{33}\lambda^3+{O}\big(\frac{\lambda^n}{\beta{^{4-n}}}\big)~.
\end{array}
\end{equation}
These include the counterterms and lattice artifacts mentioned above: 
some of the coefficients
($d_{00}$, $d_{10}$, $d_{11}$, $d_{21}$, $d_{22}$) have already
been estimated while the other ones (especially $d_{20}$ and $d_{30}$)
are what we aim at computing by means of NSPT. 
We will again measure the observable for different values of the
lattice extent $L$ and the parameters $\ln\beta$ and $y_\rmi{latt}$,
and carry out an extrapolation in $L$ (at fixed $\ln\beta$ and
$y_\rmi{latt}$) to get the infinite-volume results, from which we infer 
the behavior of the coefficients when varying the other variables.

%%%%%%%%%%%%%%%%%%%%%%%%%%%%%%%%%%%%%%%%%%%%%%%%%%%%%%%%%%%%%%%%%%%%%%%%%%%%
%

\section{Renormalization and improvement in the Electrostatic sector: 
first tests}

The statistics we have collected so far is sufficient just to check
the reliability of this approach.  A first test is to compare
our numerical estimates for the coefficient $d_{00}$ at fixed $L$ and
$y_\rmi{latt}$ with the analytical results: this comparison is shown both
in Table 2 and in Fig.~3 and appears satisfactory.
%
%%%%%%%%%%%%%%%%%%%%%%%%%%%%%%%%%%%%%%%%%%%%%%%%%%%%%%%%%%%%%%%%%%%%%%%%
%
\begin{table}[h]
\begin{center}
\begin{tabular}{|r|r|r|}
\hline
L & Exact result & NSPT estimate \\
\hline
5 & 0.6861  &    0.6867(11)\\
\hline
6 & 0.6833  &    0.6845(8)~\\
\hline
7 & 0.6825  &    0.6837(6)~\\
\hline
8 & 0.6822  &    0.6825(5)~\\
\hline
9 & 0.6821  &    0.6822(4)~\\
\hline
10 & 0.6821 &    0.6829(4)~\\
\hline
\end{tabular}
\caption{Comparison between exact and NSPT values of $d_{00}$,
for $y_\rmi{latt}=1.0$.}
\label{Tab.2}
\end{center}
\end{table}
%%%%%%%%%%%%%%%%%%%%%%%%%%%%%%%%%%%%%%%%%%%%%%%%%%%%%%%%%%%%%%%%%%%%%%%%%%%%%
 
As a second check, we inspect how our finite-volume data approach 
the infinite-volume limit at those orders for which we have a direct ``exact''
estimate of this limit: an example is given in Fig.~4 for the
coefficient $d_{21}$ (whose infinite-volume value is 1.4072). 
Once again, the behavior looks encouraging; 
the same is observed for the terms not shown here.

%%%%%%%%%%%%%%%%%%%%%%%%%%%%%%%%%%%%%%%%%%%%%%%%%%%%%%%%%%%%%%%%%%%%%%%%%%%%
%  
\begin{figure}[t]
 \hfill
 \begin{minipage}[t]{.45\textwidth}
  \begin{center} 
   \epsfig{file=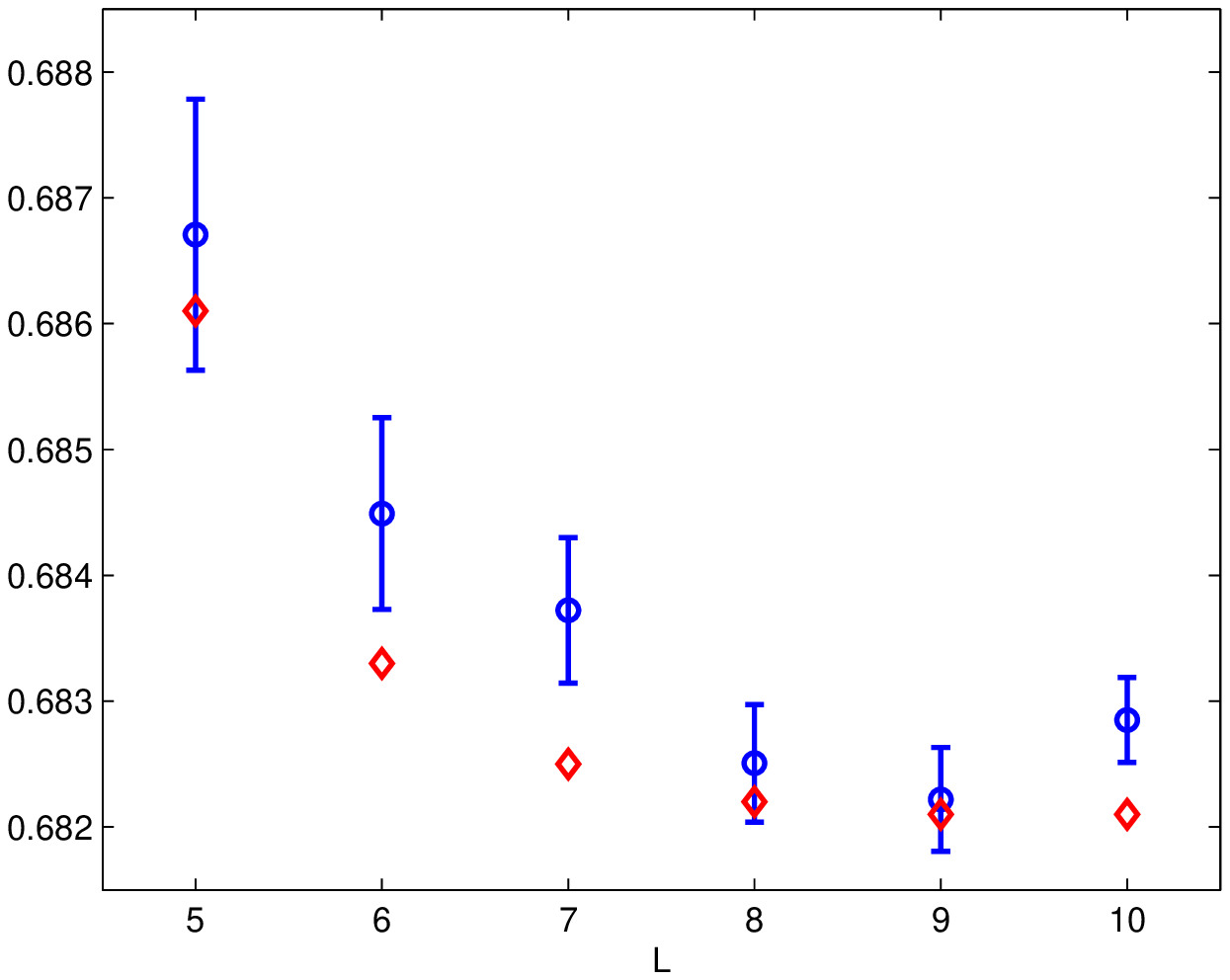, scale=0.48}
   \caption{The coefficient $d_{00}$ vs $L$;
   exact (diamonds) and NSPT (circles) results.}
   \label{Fig.3}
  \end{center}
 \end{minipage}
 \hfill
 \begin{minipage}[t]{.45\textwidth}
  \begin{center} 
   \epsfig{file=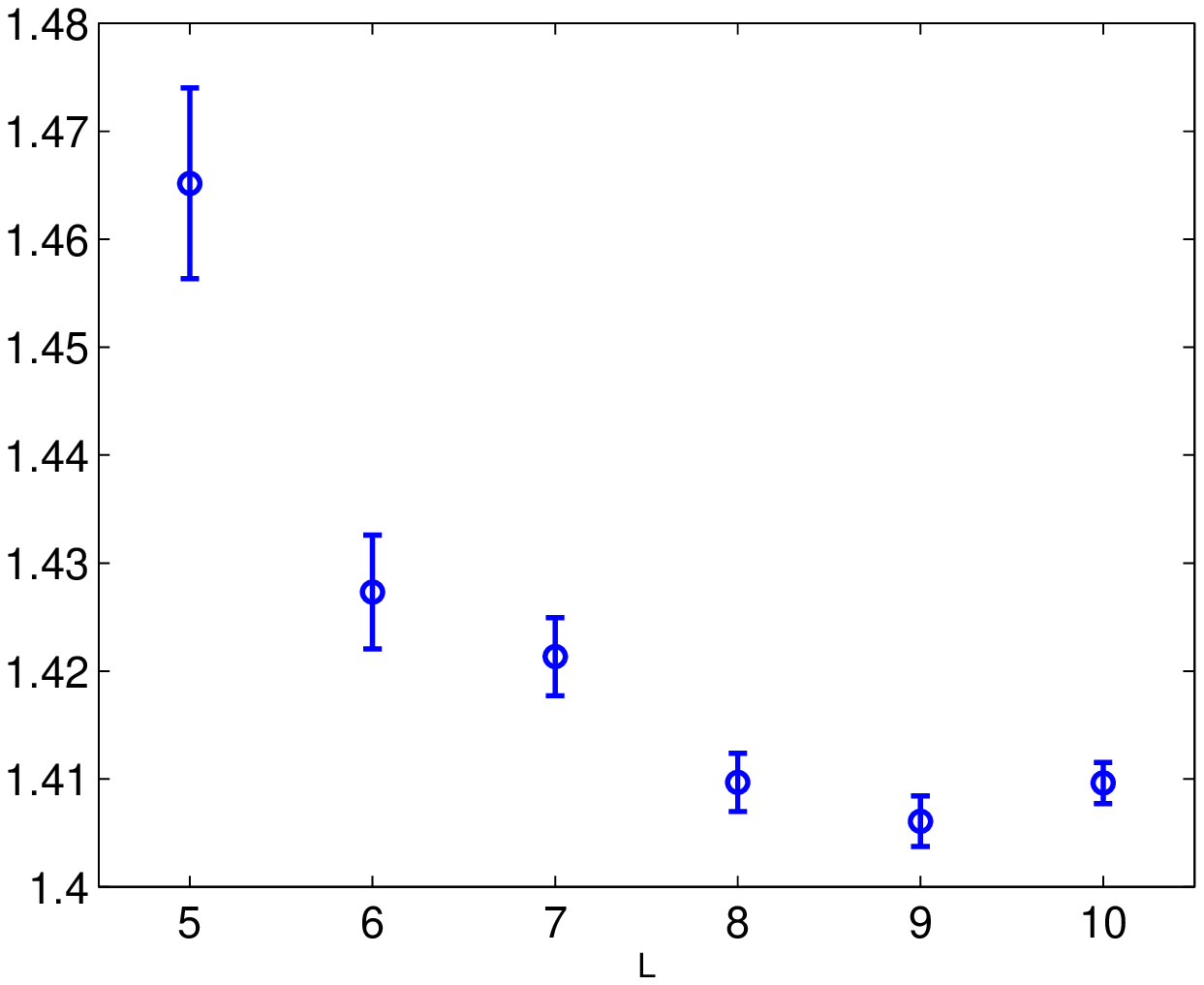, scale=0.48}
   \caption{The coefficient $d_{21}$ vs $L$.
  The expected infinite-volume value is $1.4072$.}
   \label{Fig.4}
  \end{center}
 \end{minipage}
 \hfill
\end{figure}
%
%%%%%%%%%%%%%%%%%%%%%%%%%%%%%%%%%%%%%%%%%%%%%%%%%%%%%%%%%%%%%%%%%%%%%%%%%%%

%%%%%%%%%%%%%%%%%%%%%%%%%%%%%%%%%%%%%%%%%%%%%%%%%%%%%%%%%%%%%%%%%%%%%%%%%%%%
%

\section{Conclusions and prospects}

While our determination of the renormalization constant related to 
the ${O}(g^6)$ contribution to the QCD pressure from the 
Magnetostatic sector has recently been completed~\cite{OURS}, there is 
still work to do as regards the Electrostatic contributions. It is 
important to finalise this task, since the EQCD result has 
a wider range of applicability than the MQCD result alone.
The first tests have produced encouraging results, so that 
there is every reason to believe that the determination of
the most important new coefficients ($d_{20}$ and $d_{30}$)
is also feasible, at least in a certain range of $y_\rmi{latt}$.
With these results, the program initiated in ref.~\cite{a0cond}
could finally be carried out to completion.

%%%%%%%%%%%%%%%%%%%%%%%%%%%%%%%%%%%%%% SECTION %%%%%%%%%%%%%%%%%%%%%%%%%%%%%%%
%
\section*{Acknowledgments}
%

% The Parma group acknowledges support from MIUR under 
% contract 2004023950\underline{~}002 and 
% by I.N.F.N.\ under {\em i.s.~MI11}. F.D.R.\ and Y.S.\ also acknowledge 
% support by the {\em Bruno Rossi INFN-MIT exchange program} at an early 
% stage of this project. 

We warmly thank {\em ECT*, Trento,} 
for providing computing time on the {\em BEN} system. 

%%%%%%%%%%%%%%%%%%%%%%%%%%%%%%%%%%%%%%%%%%%%%%%%%%%%%%%%%%%%%%%%%%%%%%%%%%%%
%

\end{document}